\documentclass[aps,prl,preprint,nopacs,superscriptaddress]{revtex4}
\usepackage{amsmath}
\usepackage{amssymb}
\usepackage{graphicx}
\usepackage{hyperref}
\newcommand{\beq}{\begin{equation*}}
\newcommand{\eeq}{\end{equation*}}
\newcommand{\mwt}{Mo$_x$W$_{1-x}$Te$_2$}
\newcommand{\tai}{TaIrTe$_4$}
\newcommand{\inv}{$\mathcal{I}$}
\newcommand{\tr}{$\mathcal{T}$}
\newcommand{\eb}{$E_\textrm{B}$}
\newcommand{\ef}{$E_\textrm{F}$}

\begin{document}

\title{A minimal, ``hydrogen atom'' version of an inversion-breaking Weyl semimetal}

\author{Ilya Belopolski\footnote{These authors contributed equally to this work.}} \email{ilyab@princeton.edu}
\affiliation{Laboratory for Topological Quantum Matter and Spectroscopy (B7), Department of Physics, Princeton University, Princeton, New Jersey 08544, USA}

\author{Peng Yu$^*$}
\affiliation{Centre for Programmable Materials, School of Materials Science and Engineering, Nanyang Technological University, 639798, Singapore}

\author{Daniel S. Sanchez}
\affiliation{Laboratory for Topological Quantum Matter and Spectroscopy (B7), Department of Physics, Princeton University, Princeton, New Jersey 08544, USA}

\author{Yukiaki Ishida}
\affiliation{The Institute for Solid State Physics (ISSP), University of Tokyo, Kashiwa-no-ha, Kashiwa, Chiba 277-8581, Japan}

\author{Tay-Rong Chang}
\affiliation{Department of Physics, National Tsing Hua University, Hsinchu 30013, Taiwan}

\author{Songtian S. Zhang}
\affiliation{Laboratory for Topological Quantum Matter and Spectroscopy (B7), Department of Physics, Princeton University, Princeton, New Jersey 08544, USA}

\author{Su-Yang Xu}
\affiliation{Laboratory for Topological Quantum Matter and Spectroscopy (B7), Department of Physics, Princeton University, Princeton, New Jersey 08544, USA}

\author{Daixiang Mou}
\affiliation{Ames Laboratory, U.S. DOE and Department of Physics and Astronomy, Iowa State University, Ames, Iowa 50011, USA}

\author{Hao Zheng}
\affiliation{Laboratory for Topological Quantum Matter and Spectroscopy (B7), Department of Physics, Princeton University, Princeton, New Jersey 08544, USA}



\author{Guoqing Chang}
\affiliation{Centre for Advanced 2D Materials and Graphene Research Centre, National University of Singapore, 6 Science Drive 2, 117546, Singapore}
\affiliation{Department of Physics, National University of Singapore, 2 Science Drive 3, 117546, Singapore}

\author{Guang Bian}
\affiliation{Laboratory for Topological Quantum Matter and Spectroscopy (B7), Department of Physics, Princeton University, Princeton, New Jersey 08544, USA}


%
%

\author{Horng-Tay Jeng}
\affiliation{Department of Physics, National Tsing Hua University, Hsinchu 30013, Taiwan}
\affiliation{Institute of Physics, Academia Sinica, Taipei 11529, Taiwan}

\author{Takeshi Kondo}
\affiliation{The Institute for Solid State Physics (ISSP), University of Tokyo, Kashiwa-no-ha, Kashiwa, Chiba 277-8581, Japan}

\author{Adam Kaminski}
\affiliation{Ames Laboratory, U.S. DOE and Department of Physics and Astronomy, Iowa State University, Ames, Iowa 50011, USA}

\author{Hsin Lin}
\affiliation{Centre for Advanced 2D Materials and Graphene Research Centre, National University of Singapore, 6 Science Drive 2, 117546, Singapore} \affiliation{Department of Physics, National University of Singapore, 2 Science Drive 3, 117546, Singapore}

\author{Zheng Liu}
\affiliation{Centre for Programmable Materials, School of Materials Science and Engineering, Nanyang Technological University, 639798, Singapore}
\affiliation{NOVITAS, Nanoelectronics Centre of Excellence, School of Electrical and Electronic Engineering, Nanyang Technological University, 639798, Singapore}
\affiliation{CINTRA CNRS/NTU/THALES, UMI 3288, Research Techno Plaza, 50 Nanyang Drive, Border X Block, Level 6, 637553, Singapore}

\author{Shik Shin}
\affiliation{The Institute for Solid State Physics (ISSP), University of Tokyo, Kashiwa-no-ha, Kashiwa, Chiba 277-8581, Japan}

\author{M. Zahid Hasan} \email{mzhasan@princeton.edu}
\affiliation{Laboratory for Topological Quantum Matter and Spectroscopy (B7), Department of Physics, Princeton University, Princeton, New Jersey 08544, USA}
\affiliation{Princeton Institute for Science and Technology of Materials, Princeton University, Princeton, New Jersey, 08544, USA}


\begin{abstract}
The recent explosion of research interest in Weyl semimetals has led to many proposed Weyl semimetal candidates and a few experimental observations of a Weyl semimetal in real materials. Through this experience, we have come to appreciate that typical Weyl semimetals host many Weyl points. For instance, the first Weyl semimetal observed in experiment, TaAs, hosts 24 Weyl points. Similarly, the \mwt\ series, recently under study as the first Type II Weyl semimetal, has eight Weyl points. However, it is well-understood that for a Weyl semimetal without inversion symmetry but with time-reversal symmetry, the minimum number of Weyl points is four. Realizing such a minimal Weyl semimetal is fundamentally relevant because it would offer the simplest ``hydrogen atom'' example of an inversion-breaking Weyl semimetal. At the same time, transport experiments and device applications may be simpler in a system with as few Weyl points as possible. Recently, \tai\ has been predicted to be a minimal inversion-breaking Weyl semimetal. However, crucially, the Weyl points and Fermi arcs live entirely above the Fermi level, making them inaccessible to conventional angle-resolved photoemission spectroscopy (ARPES). Here we use pump-probe ARPES to directly access the band structure above the Fermi level in \tai. We directly observe Weyl points and topological Fermi arcs, showing that \tai\ is a Weyl semimetal. We find that, in total, \tai\ has four Weyl points, providing the first example of a minimal inversion-breaking Weyl semimetal. Our results hold promise for accessing exotic transport phenomena arising in Weyl semimetals in a real material.
\end{abstract}

\date{\today}
\maketitle

%
%




A Weyl semimetal is a crystal which hosts emergent Weyl fermions as electronic quasiparticles. In an electronic band structure, these Weyl fermions correspond to accidental degeneracies, or Weyl points, between two bands \cite{Weyl, Peskin, Abrikosov, Nielsen, Volovik}. It is well-understood that Weyl points can only arise if a material breaks either spatial inversion symmetry, \inv, or time-reversal symmetry, \tr\ \cite{Murakami, Pyrochlore, Vish, Hosur}. At the same time, in a Weyl semimetal, symmetries of the system tend to produce copies of Weyl points in the Brillouin zone. As a result, typical Weyl semimetals host a proliferation of Weyl points. For instance, the first Weyl semimetals observed in experiment, TaAs and its isoelectronic cousins, have an \inv\ breaking crystal structure which gives rise to a band structure hosting $24$ Weyl points distributed throughout the bulk Brillouin zone \cite{TaAsUs, TaAsThyUs, TaAsThem, TaAsThyThem, NbAs, TaPUs, TaAsChen}. However, most of these Weyl points can be related to one another by the remaining symmetries of TaAs, namely two mirror symmetries, $C_4$ rotation symmetry and \tr. In the \mwt\ series, which has recently been under intensive theoretical and experimental study as a Weyl semimetal with strongly Lorentz-violating, or Type II, Weyl fermions, mirror symmetry and \tr\ relate subsets of the eight Weyl points \cite{AndreiNature, TayRong, Binghai, Zhijun, myothermowte, Adam1, Shuyun, Baumberger}. As another example, according to calculation, the Weyl semimetal candidate SrSi$_2$ hosts no fewer than 108 Weyl points, copied in sets of 18 by three $C_4$ rotation symmetries \cite{SrSi2}. Despite the rather large numbers of Weyl points which arise in these materials, it is known that for a Weyl semimetal which breaks \inv\ but maintains \tr, the minimal nonzero number of Weyl points allowed is 4. Realizing such a minimal Weyl semimetal is not only of fundamental interest, but is also practically important, because a system with fewer Weyl points may exhibit simpler properties in transport and be more suitable for device applications.

Recently, \tai\ was predicted to be a Weyl semimetal with only four Weyl points, the minimum allowed for an \inv\ breaking Weyl semimetal \cite{Koepernik}. It was further noted that the Weyl points are associated with strongly Lorentz-violating, or Type II, Weyl fermions, providing only the third example of a Type II Weyl semimetal after the \mwt\ series and LaAlGe, also under recent study \cite{AndreiNature, LaAlGe, RAlX}. Moreover, the Weyl points are well-separated in momentum space, with substantially larger topological Fermi arcs as a fraction of the size of the surface Brillouin zone than other known Weyl semimetals. Lastly, \tai\ has a layered crystal structure, which may make it easier to carry out transport experiments and develop device applications. All of these desirable properties have motivated considerable research interest in \tai. At the same time, one crucial challenge is that the Weyl points and topological Fermi arcs are predicted to live entirely above the Fermi level in \tai, so that they are inaccessible to conventional angle-resolved photoemission spectroscopy (ARPES). 

Here we directly observe Weyl points and topological Fermi arcs in \tai, realizing the first minimal \inv\ breaking Weyl semimetal. We first briefly reiterate a well-known theoretical argument that the minimum number of Weyl points for an \inv\ breaking Weyl semimetal is four. Then, we use pump-probe angle-resolved photoemission spectroscopy (pump-probe ARPES) to directly access the band structure of \tai\ above the Fermi level in experiment. We report the direct observation of Weyl points and topological Fermi arcs. Our results experimentally demonstrate that \tai\ has four Weyl points. We conclude that \tai\ can be viewed as a ``hydrogen atom'' Weyl semimetal, with the simplest configuration of Weyl points allowed in an \inv\ breaking system that maintains \tr.


We first reiterate well-known arguments that four is the minimum number of Weyl points allowed in an \inv\ breaking Weyl semimetal which respects \tr. A Weyl point is associated with a chiral charge, directly related to the chirality of the associated emergent Weyl fermion. It can be shown that for any given band the sum of all chiral charges in the Brillouin zone is zero. Further, under $\mathcal{T}$ a Weyl point of a given chiral charge at $k$ is mapped to another Weyl point of the same chiral charge at $-k$. This operation of $\mathcal{T}$ on a chiral charge is illustrated in Fig. \ref{Fig1}a on the blue colored Weyl point with $+1$ chiral charge. Now, if an $\mathcal{I}$ breaking Weyl semimetal has no additional symmetries which produce copies of Weyl points, then only the global requirement that total chiral charge vanish and $\mathcal{T}$ determine the minimum number of Weyl points. In the simplest case, \tr\ will produce two copies of Weyl points of chiral charge $+1$, as shown in Fig. \ref{Fig1}a. To balance these out the system must have two chiral charges of $-1$, also related by \tr. In this way, the minimum number of Weyl points in a \tr\ invariant Weyl semimetal is four. This simple scenario is realized in TaIrTe$_4$. The crystal structure of TaIrTe$_4$ is described by space group 31 ($Pmn2_1$), lattice constants $a=3.77$ \AA, $b=12.421$ \AA, and $c=13.184$ \AA, with layered crystal structure, see Fig. \ref{Fig1}b. We note that \tai\ takes the same space group as Mo$_x$W$_{1-x}$Te$_2$, has a unit cell doubled along $b$. To study how the Weyl points emerge in TaIrTe$_4$ we present the electronic band structure along various high-symmetry directions, see Brillouin zone and \textit{ab initio} calculation in Fig. \ref{Fig1}c, d. Enclosed by the rectangular box along $\Gamma$-$X$ is a crossing region between the bulk conduction and valence bands that gives rise to Weyl points. A more detailed calculation shows that the Weyl points have tilted over, or Type II, Weyl cones and that they live above the Fermi level, \ef\ at $k_z = 0$ \cite{Koepernik}. A cartoon schematic of the resulting constant energy contour at the energy of the Weyl points, $E_\textrm{B} = E_\textrm{W}$, is shown in Fig. \ref{Fig1}e for the (001) plane. Illustrated is a large electron pocket (blue colored) and smaller hole pockets (orange colored) oriented along $k_x$. The electron and hole pockets form Type II Weyl cones where they touch (red and blue marks). In this way, TaIrTe$_4$ has four Weyl points, the minimal number allowed in an $\mathcal{I}$ breaking Weyl semimetal. The overall electronic structure of \tai\ near \ef\ is similar to \mwt, but we note that the role of the electron and hole pockets is reversed in \tai\ relative to \mwt. More importantly, \tai\ hosts the minimal number of Weyl points four, while \mwt\ hosts eight, and \tai\ also hosts larger Fermi arc surface states. To study the expected Fermi arcs in TaIrTe$_4$, we present an energy-dispersion cut along a pair of projected Weyl points along $k_{y}$, Fig.\ref{Fig1}f. We clearly observe a large single Fermi arc surface state at $\sim 0.1$ eV above the Fermi level that is $\sim 0.25$ \AA$^{-1}$ long and connecting a pair of $\pm{1}$ chiral charged Weyl points along $k_y$. Thus, TaIrTe$_4$ serves as an tantalizing platform for realizing and studing the highly anticipated ``hydrogen atom'' Weyl semimetal. We also see that the Weyl points and Fermi arcs live well above the Fermi level, making them inaccessible by conventional ARPES.

Next, we use pump-probe ARPES to directly access the unoccupied band structure of \tai\ up to $E_\textrm{B} > 0.2$ eV and we find excellent agreement with calculation. In our experiment, we use a $1.48$eV pump laser pulse to excite electrons into low-lying states above the Fermi level, followed by a $5.92$eV probe laser pulse to perform photoemission \cite{IshidaMethods}. We study $E_\textrm{B}$-$k_x$ cuts near $\bar{\Gamma}$, Fig. \ref{Fig2}a-c, with key features marked by guides to eye in Fig. \ref{Fig2}d. Above the Fermi level, we see a beautiful, wispy crossing-like feature near $E_\textrm{B} \sim 0.15$ eV, labelled 1, and two electron-like bands, 2 and 3, extending out above \ef, forming the horns of a viking helmet. Below the Fermi level, we observe a general hole-like structure consisting of three bands, labelled 4-6. As we shift $k_y$ off $\bar{\Gamma}$, we find little change in the spectrum, suggesting that the band structure is rather flat along $k_y$ near $\bar{\Gamma}$. However, we can observe that band 4 moves downward in energy and becomes more intense with increasing $k_y$. We find an excellent match between our ARPES data and \textit{ab initio} calculation, Fig. \ref{Fig2}e-g. Specifically, we identify the same wispy crossing (green arrow) and top of band 4 (orange arrow). We can also track band 4 in $k_y$ in calculation and we find that the band moves down and becomes brighter as $k_y$ increases, in excellent agreement with the data. The electron-like structure of bands 2 and 3 and the hole-like structure of bands 5 and 6 are also both captured well by the calculation. Crucially, however, we notice a shift in energy between experiment and theory, showing that the sample is hole-doped by $\sim 0.05$ eV. Lastly, we plot a constant energy $k_x$-$k_y$ cut at $E_\textrm{B} = E_\textrm{F}$, where we see again that there is little dispersion along $k_y$ near $\bar{\Gamma}$, Fig. \ref{Fig2}h. We also indicate the locations of the $E_\textrm{B}$-$k_x$ cuts (blue lines). Our pump-probe ARPES measurements allow us to directly measure the electronic structure above \ef\ in \tai\ and we find excellent match with the calculation.

Now we demonstrate that \tai\ is a Weyl semimetal by directly studying the unoccupied band structure to pinpoint Weyl cones and topological Fermi arcs. We study \eb-$k_y$ cuts where, based on calculation, we fix $k_x$ near the locations of the Weyl points, $k_x = k_\textrm{W} = 0.2 \textrm{\AA}^{-1}$, see Fig. \ref{Fig3}a-c, with key features marked by guides to the eye in Fig. \ref{Fig3}d. We observe two cone features, labelled 1 and 2, connected by a weak, rather flat arc feature, labelled 3. We next track the cone and arc candidates for $k_x$ away from $k_\textrm{W}$, Fig. \ref{Fig3}e-g. We find that the cones are most pronounced at $k_x \sim k_\textrm{W}$, but fade for larger $k_x$. We also find that the arc disperses slightly upward, by $\sim 10$ meV (yellow arrow), while the band 4 disperses downward by $\sim 20$ meV. This dispersion is consistent with a topological Fermi arc, which should connect the Weyl points and sweep upward with increasing $k_x$. Based solely on our pump-probe ARPES spectra, we propose that \tai\ hosts two pairs of Weyl points of chiral charge $\pm 1$ at $k_x = \pm k_\textrm{W}$, connected by Fermi arcs. Especially when viewed on a \eb-$k_y$ cut, this configuration of Weyl cones and Fermi arcs is arguably the simplest possible in any Weyl semimetal, Fig. \ref{Fig3}h. To provide further evidence for this interpretation, we compare our results to calculation, Fig. \ref{Fig3}i-k. We can easily match the Weyl cones, the Fermi arc and an upper electron-like band, labelled 5 in Fig. \ref{Fig3}d. However, we note that from calculation we expect bands 1 and 4 to attach to form a single band, while in our data they appear to be disconnected. In addition, we do not observe the lower band labelled 6 in our calculation, which may be an artifact of our measurement. At the same time, we consistently observe the broad featureless intensity below the Fermi level in both theory and experiment. Crucially, again we find a mismatch in the Fermi level. In particular, the Weyl points are expected at $E_\textrm{B} \sim 0.1$ eV, but we find the Weyl points at $E_\textrm{B} \sim 0.05$ eV. This suggests that the sample is electron doped by $\sim 50$ meV, in contrast to our results in Fig. \ref{Fig2}, where we observe an apparent $\sim 50$ meV hole doping. We propose that the difference in doping of the two samples may arise because they were grown under slightly varying conditions in two separate batches. Lastly, we note that the $k_y$ position of the Weyl points shows excellent agreement in theory and experiment. The excellent agreement between our pump-probe ARPES spectra and \textit{ab initio} calculation confirms our observation of Weyl cones and topological Fermi arcs in \tai. We can better understand the configuration of Weyl points on a constant-energy $k_x$-$k_y$ cut at $E_\textrm{B} = E_\textrm{W}$, Fig. \ref{Fig3}l. We indicate the Weyl points (blue arrows) and the locations of the $E_\textrm{B}$-$k_x$ cuts (blue lines). We also note that our results are entirely consistent with the earlier prediction of a Weyl semimetal in \tai\ \cite{Koepernik}. Our pump-probe ARPES spectra demonstrate that \tai\ is a Weyl semimetal with four Weyl points.




We compare \tai\ with other Weyl semimetals and consider our results in the context of general topological theory. Weyl semimetals known to date in experiment host a greater number of Weyl points than \tai. In particular, the well-explored TaAs family of Weyl semimetals hosts 24 Weyl points and \mwt\ hosts eight Weyl points. We plot the configuration of Weyl points for TaAs, \mwt, and \tai, where red and blue circles denote Weyl points of opposite chirality, Fig. \ref{Fig4}a-c. It is also interesting to note that the length of the Fermi arc in \tai\ is much longer as a fraction of the Brillouin zone than that of TaAs or \mwt, which can be seen clearly in the projections of the Weyl points on the (001) surface of all three systems, Fig. \ref{Fig4}d-f. We see that our discovery of a Weyl semimetal in \tai\ provides the first example of a minimal ``hydrogen atom'' version of an \inv\ breaking Weyl semimetal. Our results suggest that \tai\ holds promise as a simpler material platform for studying the novel properties of Weyl semimetals in transport and applying them in novel devices.

\section{Methods}

Preliminary ARPES measurements were carried out using a home-built laser-based ARPES setup at the Ames Laboratory in Ames, Iowa, United States. Measurements were conducted under ultra-high vacuum and at temperatures $\leq 10$K. The angular and energy resolution was better than $0.1^{\circ}$ and $5$ meV, respectively, with photon energies from 5.77 eV to 6.67 eV \cite{KaminskiMethods}. Pump-probe ARPES measurements were carried out using a hemispherical Scienta R4000 analyzer and a mode-locked Ti:Sapphire laser system that delivered $1.48$ eV pump and $5.92$ eV probe pulses at a repetition rate of $250$ kHz \cite{IshidaMethods}. The time and energy resolution were $300$ fs and $15$ meV, respectively. The spot diameters of the pump and probe lasers at the sample were $250\ \mu$m and $85\ \mu$m, respectively. Measurements were carried out at pressures $<5 \times 10^{-11}$ Torr and temperatures $\sim8$ K.

We computed electronic structures using the projector augmented wave method \cite{PAW1,PAW2} as implemented in the VASP \cite{TransitionMetals, PlaneWaves1, PlaneWaves2} package within the generalized gradient approximation (GGA) \cite{GGA} schemes. Experimental lattice constants were used \cite{TaIrTe4Structure}. A 15 $\times$ 7 $\times$ 7 Monkhorst-Pack $k$-point mesh was used in the computations with a cutoff energy of 400 eV. The spin-orbit coupling (SOC) effects were included self-consistently. To calculate the bulk and surface electronic structures, we constructed first-principles tight-binding model Hamilton, were the tight-binding model matrix elements are calculated by projecting onto the Wannier orbitals \cite{MLWF1, MLWF2, Wannier90}, which use the VASP2WANNIER90 interface \cite{MLWF3}. We used Ta $d$, Ir $d$, and Te $p$ orbitals to construct Wannier functions and without perform the procedure for maximizing localization.

\section{Acknowledgements}

I.B. acknowledges the support of the US National Science Foundation GRFP. Y.I. is supported by the Japan Society for the Promotion of Science, KAKENHI 26800165. The ARPES measurements at Ames Lab were supported by the U.S. Department of Energy, Office of Science, Basic Energy Sciences, Materials Science and Engineering Division. Ames Laboratory is operated for the U.S. Department of Energy by Iowa State University under contract No. DE-AC02-07CH11358. This work is also financially supported by the Singapore National Research Foundation (NRF) under NRF RF Award No. NRF-RF2013-08, the start-up funding from Nanyang Technological University (M4081137.070). T.-R.C. and H.-T.J. were supported by the National Science Council, Taiwan. H.-T.J. also thanks the National Center for High-Performance Computing, Computer and Information Network Center National Taiwan University, and National Center for Theoretical Sciences, Taiwan, for technical support. H.L. acknowledges the Singapore NRF under Award No. NRF-NRFF2013-03.


\clearpage

\begin{figure}
\centering
\includegraphics[width=15cm, trim={0 0 0 0}, clip]{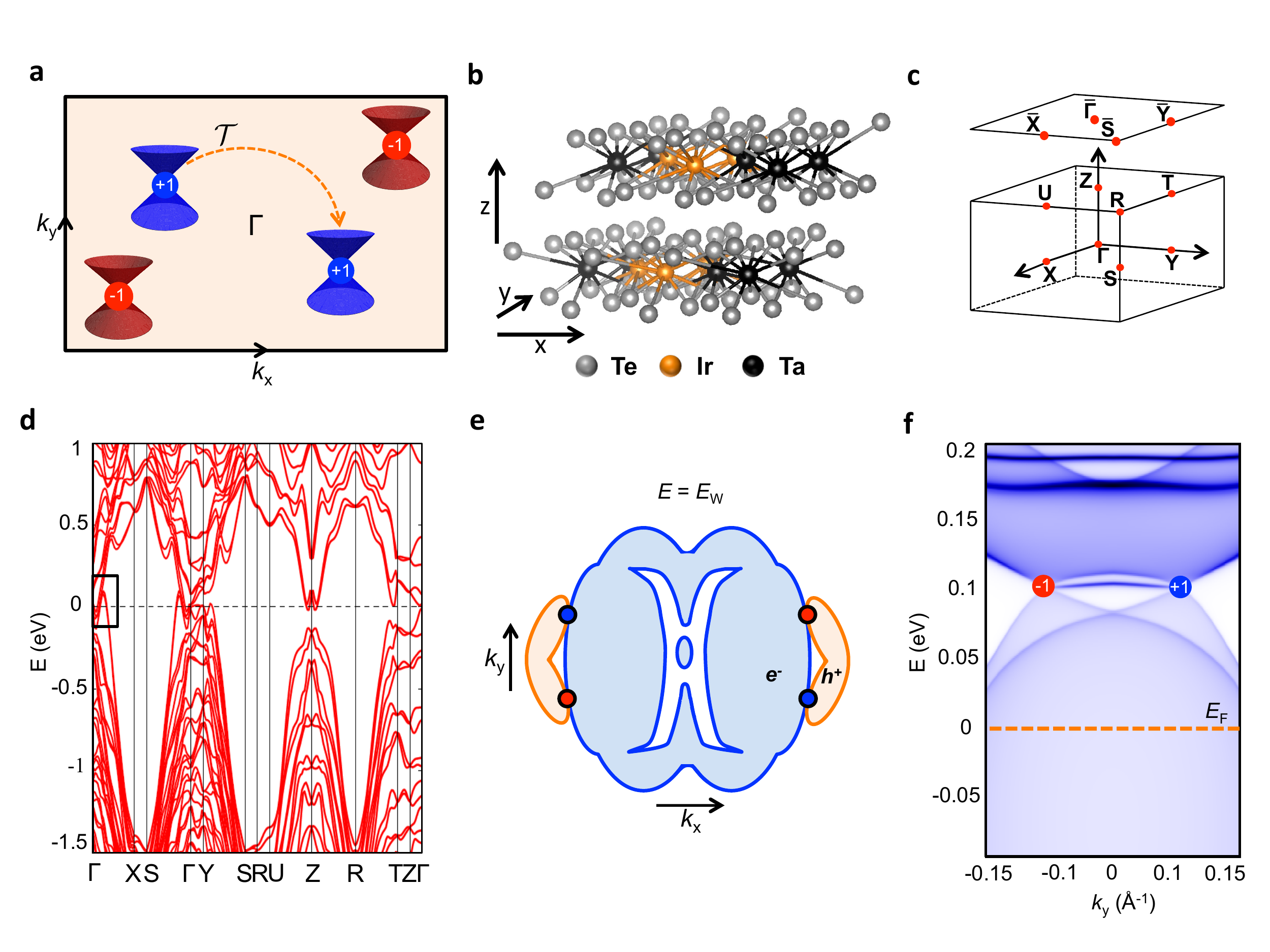}
\end{figure}
\clearpage

\begin{figure*}
\caption{\label{Fig1}\textbf{The simplest \inv\ breaking Weyl semimetal.} (a) Illustration of the minimal number of Weyl points in a $\mathcal{T}$ invariant Weyl semimetal. The blue and red circles and cones represent Weyl points and Weyl cones with $\pm{1}$ chiral charge at generic $k$-points. In a $\mathcal{T}$ invariant Weyl semimetal the minimal number of four Weyl points is generated by the operation of $\mathcal{T}$ symmetry on a single of pair oppositely chiral charged Weyl points, which preserves chiral charge (orange arrow). (b) The crystal structure of TaIrTe$_4$ is layered, in space group 31, which breaks inversion symmetry. (c) The bulk Brillouin zone (BZ) and (001) surface BZ of \tai\ with all high-symmetry points marked in red. (d) The electronic band structure of TaIrTe$_4$ along high-symmetry lines. There is a band crossing in the region near $\Gamma$. (e) Cartoon illustration of the constant energy contour at $E_{\textrm{B}}=E_{\textrm{W}}$ for the (001) surface, with bulk electron (blue) and hole (orange) pockets. A detailed calculation shows that the pockets touch to form four Type II Weyl points (blue and red circles) \cite{Koepernik}. (f) Energy-dispersion calculation along a pair of Weyl points in the $k_{y}$ direction. The Weyl points and Fermi arcs live at $\sim 0.01$ eV above $E_{\textrm{F}}$, requiring the use of pump-probe ARPES to directly access the unoccupied band structure to demonstrate a Weyl semimetal.}
\end{figure*}

\begin{figure*}
\centering
\includegraphics[width=15cm, trim={1in 1in 1in 1in}, clip]{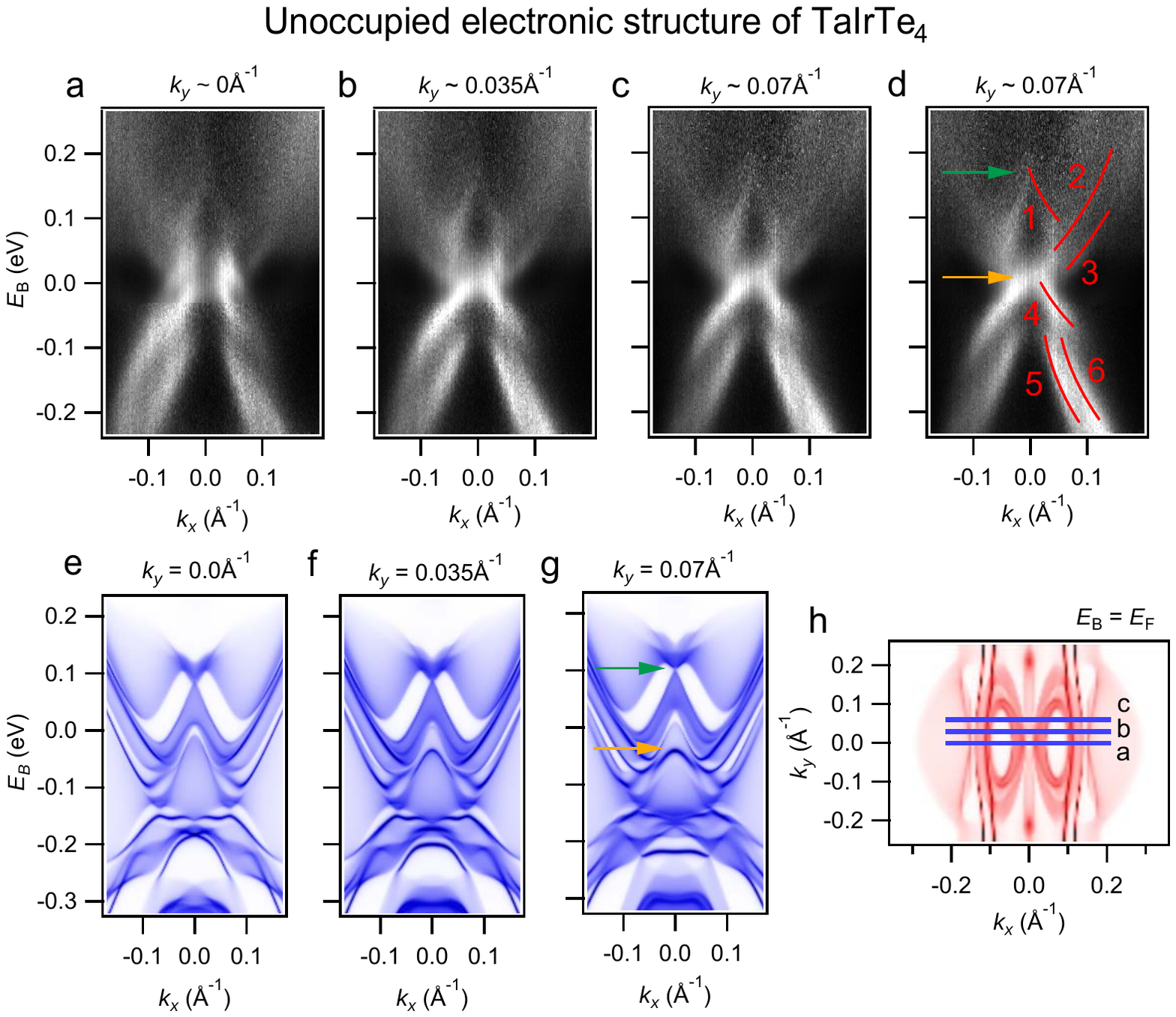}
\end{figure*}

\begin{figure*}
\caption{\label{Fig2}\textbf{Band structure of \tai\ above the Fermi level.} (a-c) Pump-probe ARPES dispersion maps of \tai, showing dispersion above \ef\ at fixed $k_y$ near $\bar{\Gamma}$. (d) Same as (c) but with key features marked. (e-f) \textit{Ab initio} calculation of \tai. The data is captured well by calculation, but the sample appears to be hole doped by $\sim 50$ meV, comparing the green and orange arrows in (d) and (g). (h) Calculation of the nominal Fermi surface, showing weak dispersion along $k_y$ near $\bar{\Gamma}$, consistent with the data. Cuts (a-c) are marked (blue lines).}
\end{figure*}
\clearpage

\begin{figure*}
\centering
\includegraphics[width=15cm, trim={1in 1in 1in 1in}, clip]{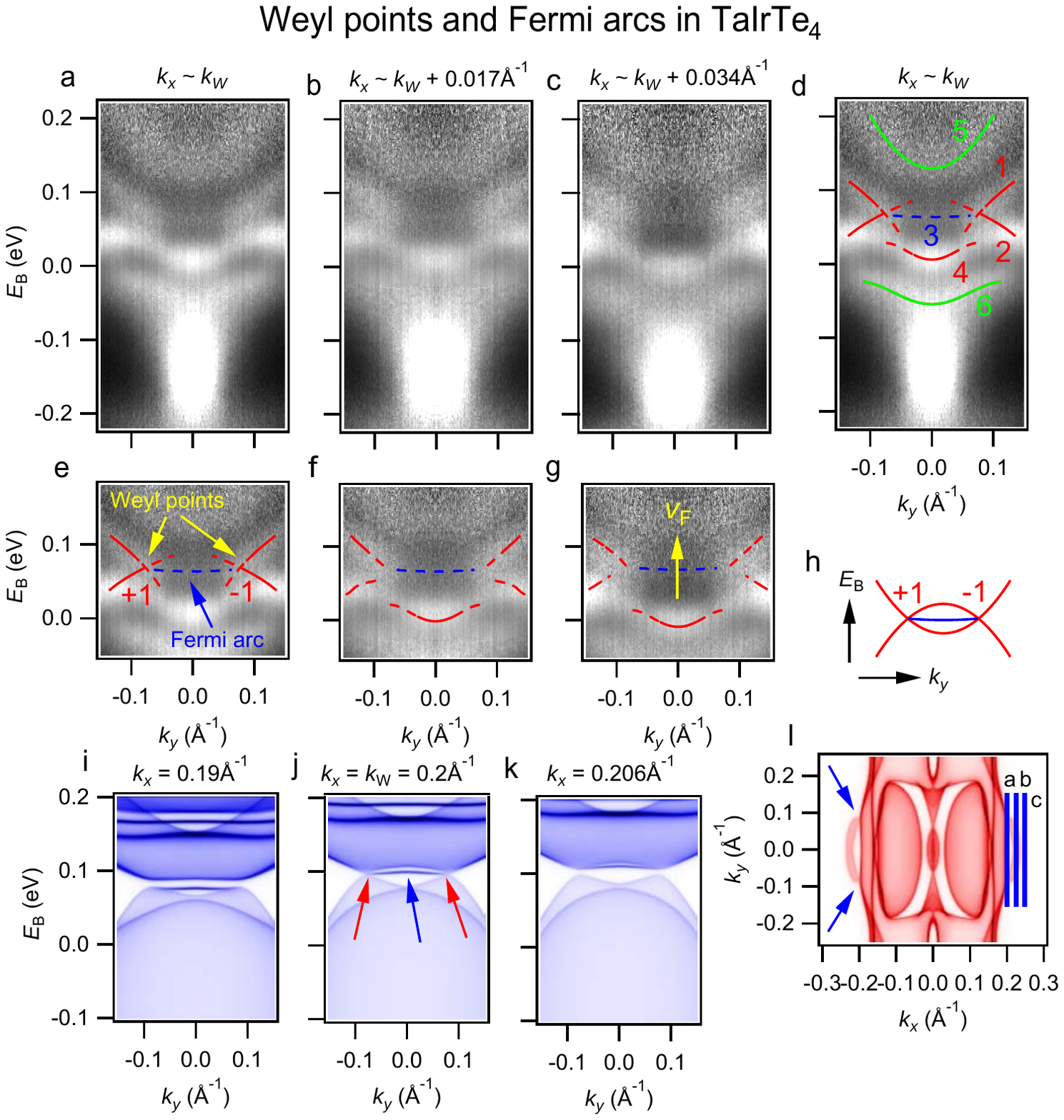}
\end{figure*}
\clearpage

\begin{figure*}
\caption{\label{Fig3}\textbf{Observation of Weyl points and Fermi arcs above the Fermi level in \tai.} (a-c) Pump probe ARPES spectra of \tai, showing dispersion above \ef\ at fixed $k_x$ expected to be near the Weyl points. (d) Same spectrum as (a) but with key features marked. The Weyl cone candidates are labelled 1 and 2, the Fermi arc candidate is labelled 3. (e-g) The same spectra as (a-c) but focusing on the Weyl points and Fermi arcs, with key features marked. We observe two cones connected by an arc, which disperses upward with increasing $k_x$, as indicated by the yellow arrow in (g) and consistent with the dispersion expected for a topological Fermi arc. The $\pm 1$ indicate the chiral charge of the Weyl points. These chiral charges are fixed to $\pm 1$ because only one Fermi arc emanates from each Weyl point. (h) Cartoon of the cones and arc observed in the data, showing what is perhaps the simplest configuration of Weyl points and Fermi arcs that can exist in any Weyl semimetal. (i-k) \textit{Ab initio} calculation of \tai\ showing the Weyl points (red arrows) and Fermi arcs (blue arrow). The excellent agreement with calculation demonstrates that we have observed a Weyl semimetal in \tai.(l) Constant-energy calculation of \tai\ at $E_\textrm{B} = E_\textrm{W}$ with Weyl points marked (blue arrows) and cuts (a-c) marked (blue lines).}
\end{figure*}
\clearpage

\begin{figure*}
\centering
\includegraphics[width=15cm, trim={60 350 80 60}, clip]{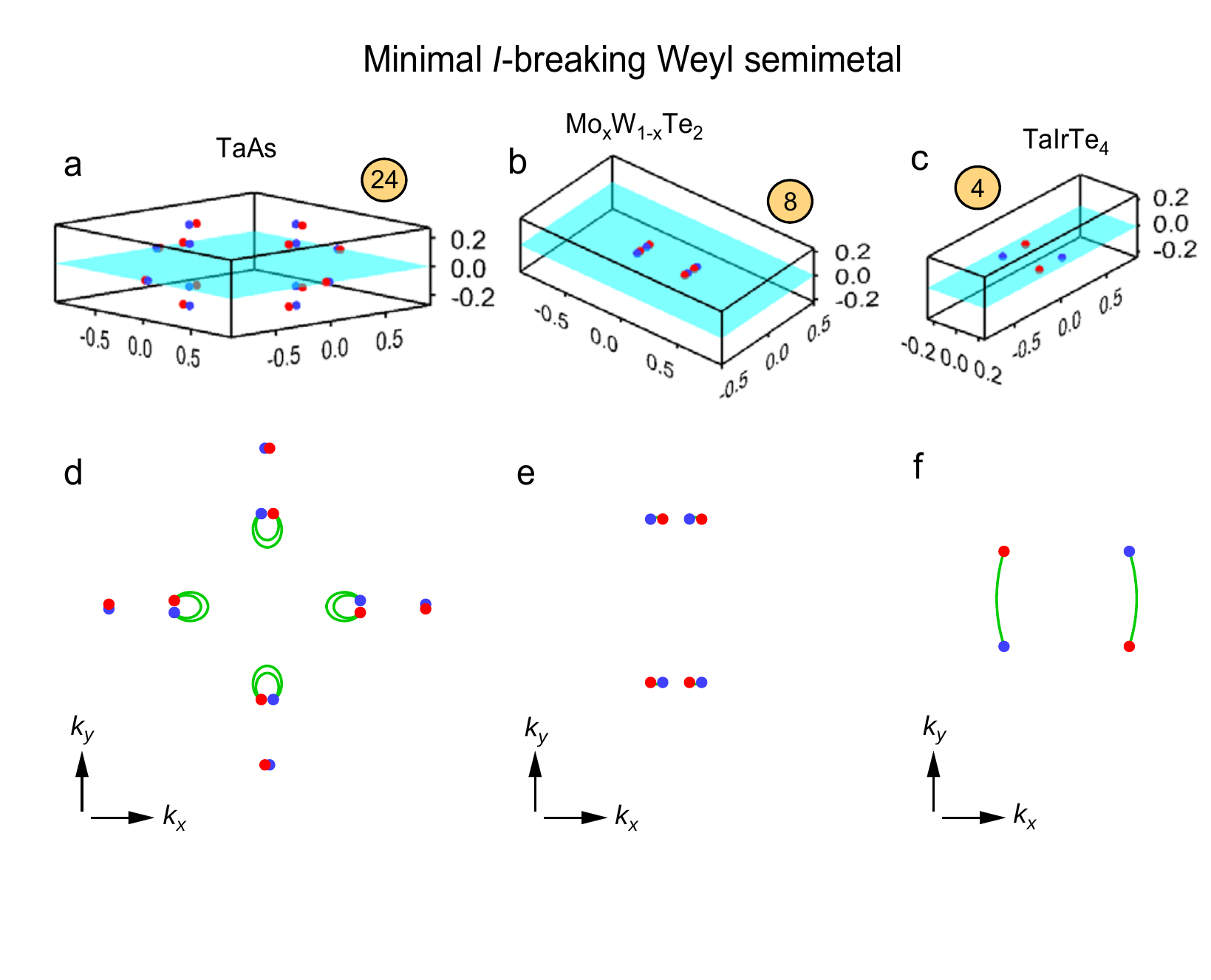}
\end{figure*}
\clearpage

\begin{figure*}
\caption{\label{Fig4}\textbf{A ``hydrogen atom'' Weyl semimetal.} Configuration of Weyl points, plotted in red and blue for Weyl points of opposite chirality, for (a) TaAs, with 24 Weyl points, (b) \mwt, with eight Weyl points and (c) \tai, with the minimal number, only four Weyl points, making \tai\ the ``hydrogen atom'' of \inv\ breaking Weyl semimetals. (d-f) The projection of the Weyl points on the (001) surface, with a cartoon of the topological Fermi arcs. The length of the Fermi arc in \tai\ is a much longer as a fraction of the Brillouin zone in comparison to the Fermi arcs in TaAs and \mwt.}
\end{figure*}
\clearpage

\end{document}